\documentstyle[preprint,eqsecnum,aps,epsfig]{revtex}
\tightenlines

\newcommand{\Sla}{\hspace{-10.5pt}~/}

\begin{document}
\draft
\preprint{HUPD-9906}
\title{Path-Integral Formulation of Casimir Effects \\
       in Supersymmetric Quantum Electrodynamics}

\author{Hiroyuki Abe, Junya Hashida, Taizo Muta and 
Agus Purwanto}
\address{Department of Physics, Hiroshima University \\
Higashi-Hiroshima, Hiroshima 739-8526, Japan}
\date{\today}
\maketitle
\begin{abstract}
The path-integral method of calculating the Casimir energy
between two parallel conducting plates is developed within 
the framework of supersymmetric quantum electrodynamics at 
vanishing temperature as well as at finite temperature.
The choice of the suitable boundary condition for the photino 
on the plates is argued and the physically acceptable condition 
is adopted which eventually breaks the supersymmetry. The 
photino mass term is introduced in the Lagrangian and the photino 
mass dependence of the Casimir energy and pressure is fully
investigated.
\end{abstract}


\narrowtext
\section{Introduction}
\noindent
The Casimir effect is an interesting phenomenon in the sense that it
provides us with one of the primitive means of extracting the energy
out of the vacuum. Since the original work of Casimir\cite{Cas,Pol} 
a number of works have appeared in extending the result to the case of
more general topological and dynamical configurations of the boundary
condition and to the circumstances at finite temperature and
gravity \cite{Book}. In the studies of the Casimir effects it is common
to assume the free electromagnetic field in the bounded region.
It may be interesting to extend our arguments for fields other than
the electromagnetic field. The Casimir effect due to the free fermionic
fields has been investigated by several authors and has been found to
result in an attractive force under the suitable physical boundary
conditions \cite{John,Gun}.

The supersymmetry is considered to be promising in searching for the
unified theory of elementary particles. In this connection it is natural
to extend the quantum electrodynamics to incorporate the supersymmetry.
It may then be interesting to study the Casimir effect in the 
supersymmetric
quantum electrodynamics in the hope that some evidence of supersymmetry
could be observed through the Casimir effect.
In the present communication we deal with the Casimir effect by applying
the path-integral formulation \cite{Gang} working in the
supersymmetric quantum electrodynamics \cite{Iga,Nak}.
In the supersymmetric quantum electrodynamics we have extra particles,
photinos (fermions), as superpartners of photons (bosons) and we have
to discuss the contributions of photinos as well as photons in the
arguments of Casimir effects.

\section{Path-Integral Formulation}
\noindent
We start with the Lagrangian for the supersymmetric quantum 
electrodynamics,

\begin{equation}
{\mathcal{L}} = {\mathcal{L}}_B+{\mathcal{L}}_F,
\end{equation}
with
\begin{equation}
{\mathcal{L}}_B = - \frac{1}{4}F_{\mu \nu}F^{\mu \nu}-
  \frac{1}{2\alpha}({\partial}^{\mu}A_{\mu})^2-i{\bar \eta}
  {\partial}^{\mu}{\partial}_{\mu}{\eta},\label{LB}
\end{equation}
and
\begin{equation}
{\mathcal{L}}_F = -i\lambda{\sigma}^{\mu}{\partial}_{\mu}{\bar \lambda},
\end{equation}
where $F_{\mu\nu}=\partial_{\mu}A_{\nu}-\partial_{\nu}A_{\mu}$ with $A_
{\mu}$
the photon field, the gauge fixing term as well as the Fadeev-Popov term
with Fadeev-Popov ghost $\eta$ are included in the Lagrangian ${\mathcal{L}}_B$
and $\lambda$ in ${\mathcal{L}}_F$ denotes the two-component photino 
field.
We adopt the covariant gauge with gauge parameter $\alpha$.
The Fadeev-Popov ghost is usually neglected in quantum electrodynamics
since it is irrelevant as far as we consider processes without photon 
loops
and in the normal physical QED processes no photon loop appears.
In the Casimir effect, however, we calculate the contribution of the 
photon loop
and so we need to take into account the Fadeev-Popov ghost
in order to count correctly the physical degrees of freedom of photons.

We first deal with the Casimir effect at vanishing temperature.
Consider the generating functional $Z$ given by
\begin{equation}
Z = \int {[dA_{\mu}][d\eta][d\bar\eta][d\lambda]}{[d\bar\lambda]}
e^{i\int d^4x {\mathcal{L}}}
\end{equation}
The free energy density $\epsilon$ is related to the generating 
functional $Z$
through the relation
\begin{equation}
Z = e^{-i\epsilon \Omega}, \label{epsdef}
\end{equation}
where $\Omega$ is the space-time volume.

\subsection{Photon}
\noindent
The contribution of photons to the generating functional $Z$ is given by
\begin{equation}
Z_B = \int {[dA_{\mu}]}{[d \eta]}{[d \bar \eta]}e^{i\int d^4x
  {\mathcal{L}}_B}.
\end{equation}
Performing integration over $A_{\mu}$, $\eta$ and $\bar\eta$ we obtain
\begin{equation}
Z_B = \left[ \textrm{Det} \left( 2i\partial^2 \right) \right]
      \left[ \textrm{Det} (-g_{\mu\nu}\partial^2 + ( 1-1/\alpha)
      \partial_{\mu}\partial_{\nu} ) \right] ^{-1/2},
\end{equation}
where the overall factor of the numerical constant is neglected.
We note that in momentum space,
\begin{equation}
\det \left[g_{\mu\nu}k^2 + \left(\frac{1}{\alpha}-1\right)
k_{\mu}k_{\nu}\right] = \frac{1}{\alpha}(k_0^2-k_1^2-k_2^2-k_3^2)^4,
\end{equation}
and hence
\begin{equation}
\frac{\ln Z_B}{\Omega} = -2 \int \frac{d^4k}{{(2\pi)}^4} ln{(-k^2)}+
  \frac{1}{2} \int \frac{d^4k}{{(2\pi)}^4}(ln \alpha -\frac{i\pi}{2})
  + \int \frac{d^4k}{{(2\pi)}^4} ln{(-k^2)}. \label{ZB}
\end{equation}
Here in Eq. (\ref{ZB}) the first, second and third term
come from the $F_{\mu\nu}$ term, the gauge term and the Fadeev-Popov 
term in the Lagrangian (\ref{LB}) respectively.
Since the second term is a constant which is physically irrelevant,
we neglect it in the following arguments.
We make the Wick rotation in Eq. (\ref{ZB}) and find
\begin{equation}
\frac{\ln Z_B}{\Omega} = -i\int \frac{d^4k}{{(2\pi)}^4}\ln k^2,
\end{equation}
for Euclidean momentum $k$.
According to the relation (\ref{epsdef}) the free energy density $\epsilon$
is then easily obtained by performing the integration over the fourth (time)
component of the Euclidean four-momentum $k_{\mu}$,
\begin{equation}
\epsilon_B = \int {\frac{d^3k}{{(2\pi)}^3} \sqrt{{\bf k}^2}},
\label{epsB}
\end{equation}
where we have neglected the additive constant stemming from
the $k_4$ integration.
Of course the free energy density $\epsilon_B$ is a divergent quantity 
if one considers the infinite space region. In the present argument of
the Casimir energy we consider the region bounded by two conducting 
plates.
We place two infinite conducting plates perpendicular to the $z$ 
direction
at $z=0$ and at $z=L$.
For photons we introduce the physical boundary condition \cite{Cas,Pol}
\begin{equation}
{\bf A}(z=0)={\bf A}(z=L)=0. \label{bcphoton}
\end{equation}
According to the condition (\ref{bcphoton}) the $z$-component $k_z$ of 
momentum ${\bf k}$ is discretized such that
\begin{equation}
k_z = \frac{\pi n}{L},
\end{equation}
with $n$ an integer.
Replacing the $k_z$ integration by summation on $n$ in Eq. (\ref{epsB})
we obtain
\begin{equation}
\epsilon_B = \frac{1}{2L}\sum_{n=-\infty }^{\infty}\int \frac{d^2k}{{(2\pi)}^2}
             \sqrt{{{\bf k}_T}^2+{(\pi n/L)}^2}, \label{epsBB}
\end{equation}
where ${\bf k}_T$ refers to the transversal component of ${\bf k}$,
i. e., $(k_x, k_y)$.
The free energy density (\ref{epsBB}) is a divergent quantity and we
regularize it by introducing the regularization factor
$e^{-\tau \omega}$ in the integration (\ref{epsBB}) with
$\omega=\sqrt{{{\bf k}_T}^2+{(\pi n/L)}^2}$,
\begin{equation}
\epsilon_B = \frac{1}{L}\sum_{n=0}^{\infty}\int \frac{d^2k}{{(2\pi)}^2}
             \omega e^{-\tau \omega}. \label{epsBBR}
\end{equation}
The integration and summation in Eq. (\ref{epsBBR}) can be easily 
performed
resulting in
\begin{equation}
\epsilon_B = \frac{3}{\pi^2 \tau^4}- \frac{{\pi}^2}{720L^4}.
\end{equation}
We distinguish the free energy density for the infinite volume (Eq. (\ref{epsB}))
from the one for the bounded region (Eq. (\ref{epsBBR})) by denoting the 
former
by ${\epsilon_B}^\infty $. If we use the same regularization as
in Eq. (\ref{epsBBR}),
we obtain ${\epsilon_B}^\infty = 3/({\pi}^2{\tau}^4)$. The Casimir 
energy
${\epsilon}_{CB}$ then is given by
\begin{equation}
{\epsilon}_{CB}=\epsilon_B-{\epsilon_B}^\infty = -\frac{\pi^2}{720L^4}.
\end{equation}

\subsection{Photino}
\noindent
The contribution of photinos to the generating functional $Z$ is given by
\begin{equation}
Z_F = \int {[d \lambda]}{[d \bar \lambda]}e^{i\int d^4x {\mathcal{L}}_F}.
\end{equation}
After integrating over the photino fields we have
 $Z_F=\textrm{const.}\times \textrm{Det}(i \partial \Sla)$.
Going over to the momentum representation and performing the Wick 
rotation we find
\begin{equation}
\frac{\ln Z_F}{\Omega} = i \int{\frac{d^4k}{(2\pi)^4}\ln k^2}.
\end{equation}
Then the photino contribution to the free energy density is given by
\begin{equation}
\epsilon_F = -\int {\frac{d^3k}{{(2\pi)}^3} \sqrt{{\bf k}^2}}. \label{epsF}
\end{equation}
For the infinite volume Eq. (\ref{epsF}) gives the divergent expression
\begin{equation}
{\epsilon_F}^\infty = -\frac{3}{\pi^2 \tau^4},
\end{equation}
with $\tau$ the same regularization parameter as in ${\epsilon_B}^\infty$.

Consider the bounded region where two parallel plates are placed 
perpendicular to the $z$ axis at $z=0$ and $z=L$ respectively.
The boundary condition for photino
fields should be chosen by physical requirements. The physically 
acceptable requirement may be that there is no fermionic outgoing flux 
perpendicular to the boundaries. This requirement has been adopted to
the MIT bag model long time ago \cite{John,Chod}. It is satisfied by
the following condition at the boundary plates \cite{Gun},
\begin{equation}
i n\Sla \lambda = \lambda,\label{bcphotino}
\end{equation}
where $n_\mu=(0, \bf n)$ with $\bf n$ the inward normal to the boundary 
plates.
According to the condition (\ref{bcphotino}) the $z$-component $k_z$ of 
momentum ${\bf k}$ is discretized such that
\begin{equation}
k_z = \frac{\pi (2n+1)}{2L},
\end{equation}
with $n$ an integer.

It should be noted here that the supersymmetry under consideration is 
explicitly
broken according to the fact that the boundary condition for photinos is
different from that for photons. If the supersymmetry is to be respected
even in the bounded region, then of course one has to choose a specific
(but rather unphysical) boundary condition for photinos in order to 
maintain the supersymmetry. In our investigation we rather choose the
physically acceptable
boundary condition for photinos and consider that the supersymmetry
is an exact symmetry only for the unbounded space-time.
The situation is in some similarity with the finite temperature
case where the supersymmetry is broken explicitly according to
the different boundary conditions for bosons and fermions respectively.

Replacing the $k_z$ integration by summation on $n$ in Eq. (\ref{epsF})
we obtain
\begin{equation}
\epsilon_F = -\frac{1}{2L}\sum_{n=-\infty }^{\infty}\int \frac{d^2k}{{(2\pi)}^2}
             \sqrt{{{\bf k}_T}^2+{\{ \pi(2n+1)/2L \} }^2}. \label{epsFB}
\end{equation}
The free energy density (\ref{epsFB}) is also divergent and we 
regularize it
by using the same regularization factor as in Eq. (\ref{epsBBR}). 
Performing the
summation and integration we obtain
\begin{equation}
\epsilon_F = -\frac{3}{\pi^2 \tau^4}- \frac{7}{8}\frac{{\pi}^2}{720L^4}.
\label{epsFBR}
\end{equation}
The Casimir energy due to the fermionic effect is given by
\begin{equation}
{\epsilon}_{CF}=\epsilon_F-{\epsilon_F}^\infty = -\frac{7}{8}\frac{\pi^2}
{720L^4}.
\end{equation}

The total Casimir energy for the supersymmetric quantum electrodynamics 
is
given by summing up the photon and photino contributions and reads
\begin{equation}
{\epsilon}_C={\epsilon}_{CB}+{\epsilon}_{CF} = -\frac{15}{8}\frac{\pi^2}
{720L^4}.
\end{equation}
It is interesting to note here that the divergences appearing in ${\epsilon}_B$
and ${\epsilon}_F$ cancel out when they are added up and thus
\begin{equation}
{\epsilon}_C={\epsilon}_B+{\epsilon}_F.
\end{equation}

\section{Temperature Dependence}
\noindent
We now introduce the temperature. The Matsubara formalism \cite{Mat} 
will be employed to introduce temperature effects in the following
calculation.
We replace the integral in the energy variable $k_0$ by the summation 
where we apply the periodic boundary condition for boson fields and
the anti-periodic boundary condition for fermion fields respectively:
\begin{equation}
\int\frac{dk_0}{2\pi}f(k_0)\rightarrow\frac{1}{\beta}
\sum_{n=-\infty}^{\infty}f(\omega_n),
\end{equation}
where $\beta=1/T$ with $T$ the temperature and
the energy variable $k_0$ in the summation is replaced
by $\omega_n$ which is given by 
\[
\omega_n = \left\{
\begin{array}{ll}
        \omega_n^B\equiv\frac{2n}{\beta}\pi&(\textrm{boson}),\\
\\
        \omega_n^F\equiv\frac{2n+1}{\beta}\pi&(\textrm{fermion}).
\end{array}
\right.
\]
Repeating the similar calculations as in the previous zero-temperature
case we obtain the temperature dependent Casimir energy for photons
and photinos respectively:
\begin{eqnarray}
\epsilon_B(T) = \frac{1}{2L}\sum_{n=-\infty }^\infty 
  [\int \frac{d^2k}{{(2\pi)}^2}
  \sqrt{k^2}+\frac{2}{\beta}\int \frac{d^2k}{{(2\pi)}^2}
  ln(1-e^{-\beta\omega_B})],\label{epsBT} \\
\epsilon_F(T) = \frac{1}{2L}\sum_{n=-\infty }^\infty 
  [-\int \frac{d^2k}{{(2\pi)}^2}
  \sqrt{k^2}+\frac{2}{\beta}\int \frac{d^2k}{{(2\pi)}^2}
  ln(1+e^{-\beta\omega_F})],\label{epsFT}
\end{eqnarray}
where $\omega_B$ and $\omega_F$ are defined by
\begin{equation}
\omega_B = \sqrt{{{\bf k}_T}^2+{(\pi n/L)}^2}, \ \ \
\omega_F = \sqrt{{{\bf k}_T}^2+{(\pi (2n+1)/2L)}^2}.
\end{equation}
Performing the integrations in Eqs. (\ref{epsBT}) and (\ref{epsFT})
with the same regularization factors as before we obtain for photons,
\begin{eqnarray}
\epsilon_B(T) &=& \frac{3}{\pi^2 \tau^4}- \frac{{\pi}^2}{720L^4} \\ \nonumber & &
    - \sum_{n=1}^\infty     \left( \frac{T^2}{n^2 \pi} \right)^2 
    \left\{ \frac{n \pi}{2\xi} \coth \left( \frac{n \pi}{2\xi} \right)
    + \left( \frac{n \pi}{2 \xi} \right)^2 
    \textrm{cosech}^2 \left( \frac{n \pi}{2 \xi} \right) \right\},
    \label{epsBRT}
\end{eqnarray}
and for photinos,
\begin{eqnarray}
\epsilon_F(T) &=& -\frac{3}{\pi^2 \tau^4}
    - \frac{7}{8}\frac{{\pi}^2}{720L^4} \\ \nonumber & &
    + \frac{1}{4 \pi L^4}\sum_{l=1}^{\infty} (-1)^l 
    \frac{\xi^2}{l^2 \sinh (l \pi/2 \xi)} \left\{ \frac{2 \xi}{l} 
    + \pi \coth \left( \frac{l \pi}{2 \xi} \right) \right\}.
    \label{epsFRT}
\end{eqnarray}
where $\xi = L/\beta = LT$ and $\tau$ is the cut-off parameter.
The total contribution of photons and photinos in the supersymmetric
quantum electrodynamics to the Casimir energy is given by
$\epsilon_C(T)=\epsilon_B(T)+\epsilon_F(T)$. We have shown the behaviors of
$\epsilon_C(T)$ as a function of $T$ for $L$ fixed in Fig. 1. Also shown in
Fig. 1 are the individual contribution of photons and photinos to the
Casimir energy $\epsilon_{CB}(T)$ and $\epsilon_{CF}(T)$ where the divergent
parts in Eqs. (\ref{epsBRT}) and (\ref{epsFRT}) are eliminated as
in the case of the vanishing temperature.
Note here that for low temperature $\xi \ll 1$
\begin{equation}
\epsilon_B(T) = \frac{3}{\pi^2 \tau^4}- \frac{{\pi}^2}{720L^4}
  -\frac{{\xi}^2}{2L^4}\left\{1+ \frac{2\xi}{\pi}\right\}
   e^{-\pi / 2\xi},
\end{equation}
and for high temperature $\xi \gg 1$
\begin{equation}
\epsilon_B(T) = \frac{3}{\pi^2 \tau^4}- \frac{{\pi}^2}{720L^4}
-\frac{7}{8} \frac{{\pi}^2}{45}T^4
\end{equation}

\section{Photino Mass Dependence}
\noindent
In our calculations we have kept the photino massless in order to
respect the supersymmetry at vanishing temperature. Experimentally,
however, the massless photino has not been observed. One of the
possibilities of explaining the situation may be that the photino could
be extremely massive and is not be observed in the low energy experiments.
The photino mass would be generated by some supersymmetry breaking mechanism.
We consider this possibility and recalculate the Casimir energy with
massive photinos.
The calculation is straightforward and the result reads as follows:
\begin{eqnarray}
\epsilon_F(T,m) = - \frac{1}{4 \pi L} \sum_{n=-\infty}^{\infty} 
    \left\{ \frac{{\alpha_n}^2}{\tau} + \frac{2 \alpha_n}{\tau^2} 
    + \frac{2}{\tau^3} \right\} e^{- \tau \alpha_n}\\ \nonumber
    + \frac{1}{2 \pi L} \sum_{n=-\infty}^{\infty} \sum_{l=1}^{\infty}
    (-1)^l \left( \frac{T^3}{l^3} 
    + \frac{T^2}{l^2} \alpha_n \right) e^{-l {\alpha_n}/T},
\end{eqnarray}
where $\alpha_n = \sqrt{m^2 + \left( \frac{2n+1}{2L} \pi \right)^2}$ with
m the photino mass.
The behavior of the Casimir energy $\epsilon_C$ at vanishing temperature
with massive photinos is given in Fig. 2 as a function of the photino
mass $m$ together with the individual contributions of photons and photinos
respectively.

Finally we calculate the Casimir pressure by differentiating the energy 
$\epsilon_C(T,m) = \epsilon_{B}(T)+\epsilon_{F}(T,m)$ in terms of
distance $L$,
\begin{equation}
p(T,m) = - \frac{\partial}{\partial L}[\epsilon_C(T,m) L]\Big|_T.
\end{equation}
We have shown the behaviors of $p(T,m)L^4$ as a function of 
$T$ and $m$ in FIG. 3 and 4 respectively.
It may be more transparent to show the Casimir energy and
the Casimir pressure in the 2 dimensional plot with regards to
$T$ and $m$ and those are given in FIG. 5 and 6 respectively.

Judging from the rather strong dependence of the Casimir energy
as well as the Casimir pressure on the photino mass as shown
in FIG. 5 and 6, it may be rather difficult to detect the
evidence of the supersymmetry in the presently accessible
experimental situations.
We hope that future improvements of the experimental situation
will resolve the difficulty.

\acknowledgments

\noindent
The authors would like to thank Masahito Ueda for useful discussions.
One of the authors (T. M.) is indebted to Grant-in-Aid for
Scientific Research (C) provided by the Ministry of Education,
Science, Sports and Culture for the financial support under
the contract number 08640377.

\pagebreak[0]

\begin{figure}[htbp]
   \centerline{\epsfig{figure=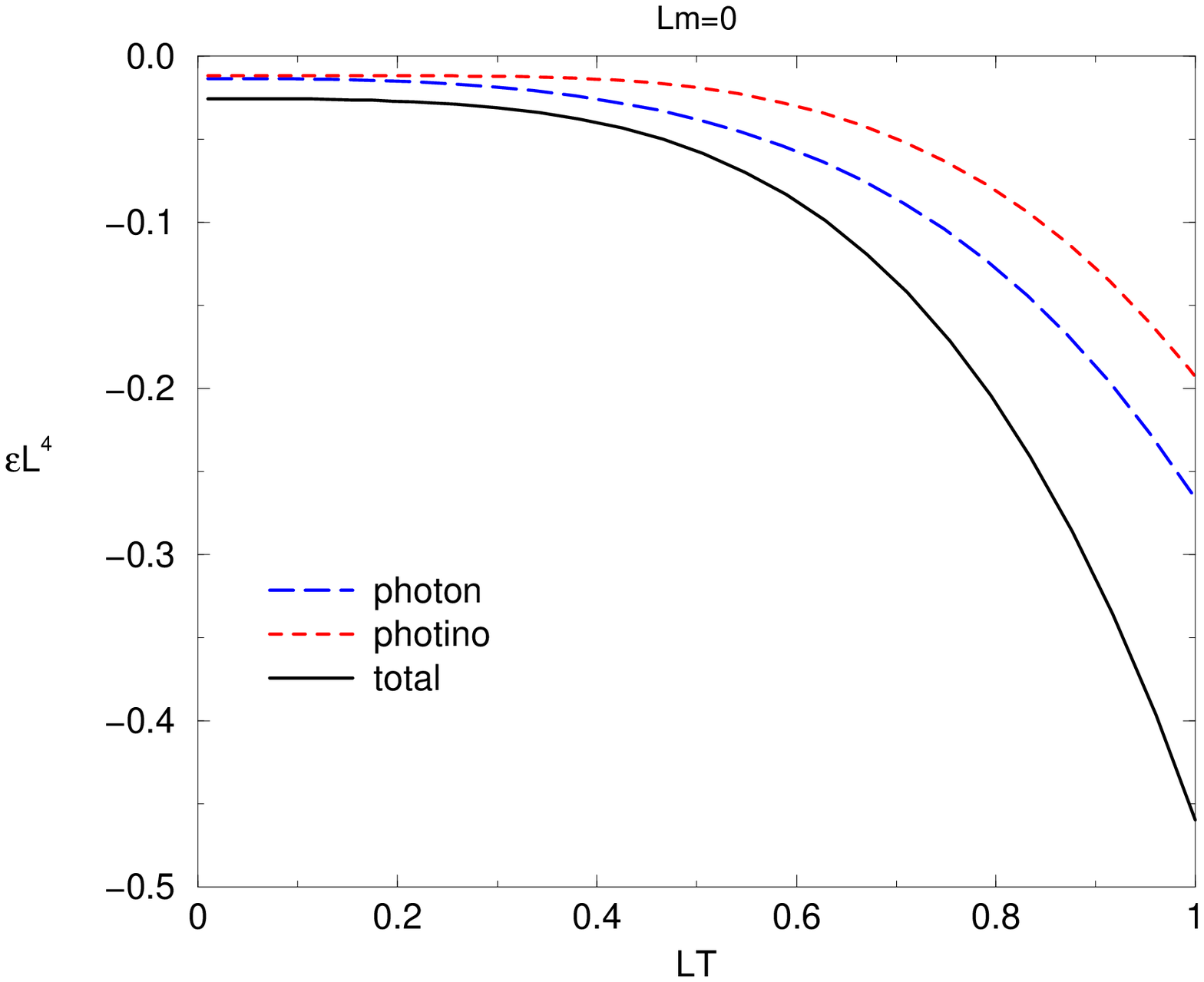,width=0.65\linewidth}}
   \caption{The behavior of Casimir energy $\epsilon_C L^4$ 
   as a function of temperature $LT$. Also shown 
   are $\epsilon_{CB} L^4$ and $\epsilon_{CF} L^4$ respectively}
\end{figure}
\begin{figure}[htbp]
   \centerline{\epsfig{figure=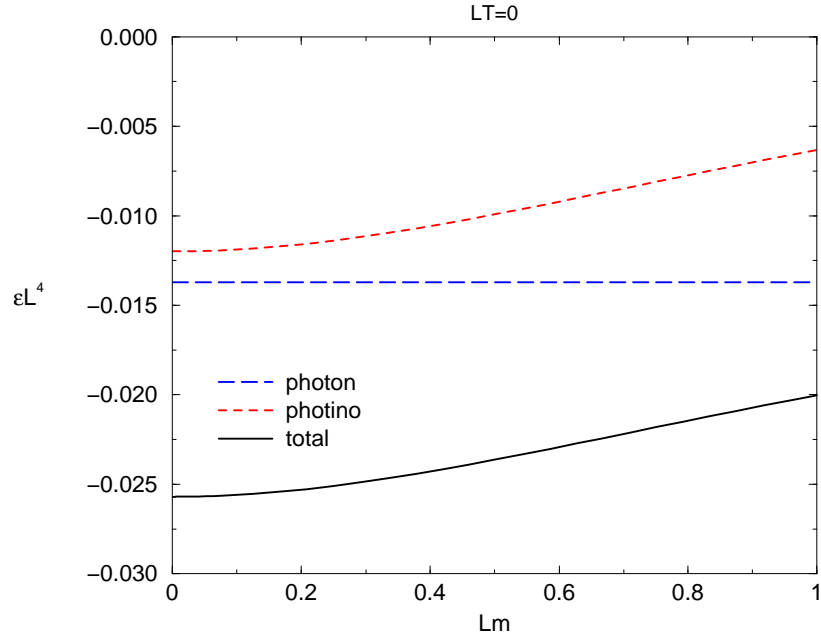,width=0.65\linewidth}}
   \caption{The behavior of Casimir energy $\epsilon_C(T) L^4$ 
   as a function of photino mass $Lm$. Individual
   contributions $\epsilon_{CB}(T) L^4$ and $\epsilon_{CF}(T) L^4$ 
   respectively are also shown.}
\end{figure}
\begin{figure}[htbp]
   \centerline{\epsfig{figure=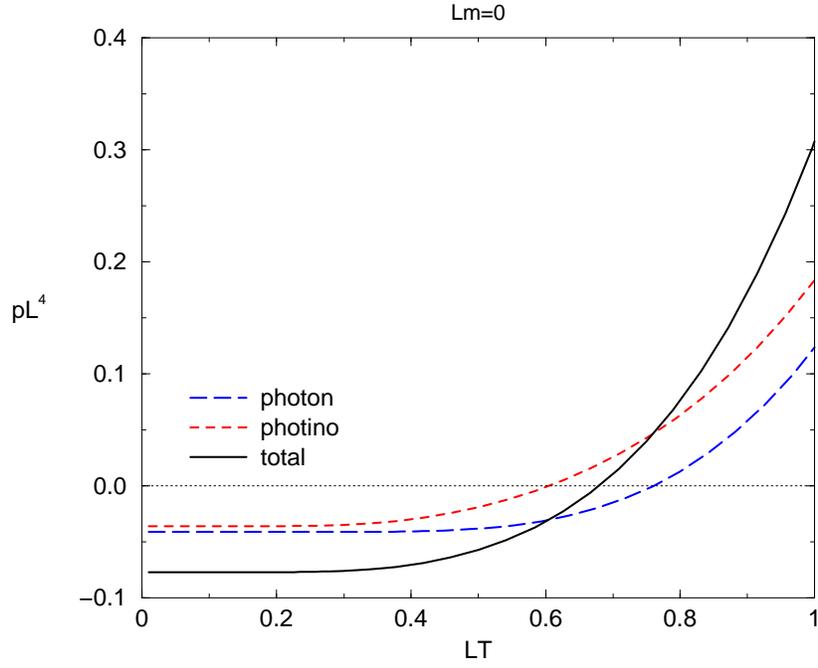,width=0.65\linewidth}}
   \caption{The behaviors of Casimir pressure $p L^4$ as a function of
   temperature $LT$.}
\end{figure}
\begin{figure}[htbp]
   \centerline{\epsfig{figure=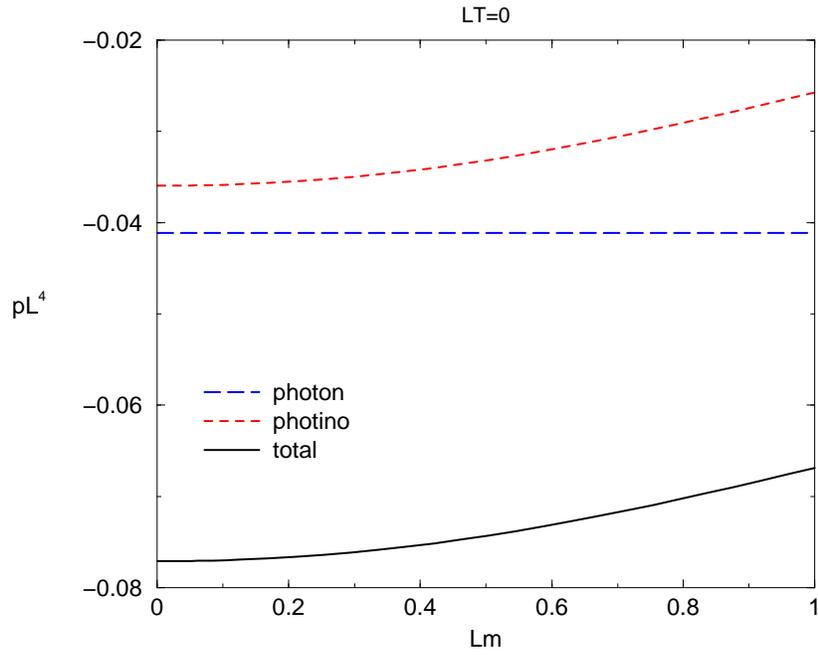,width=0.65\linewidth}}
   \caption{The behaviors of Casimir pressure $p L^4$ as a function of
   photino mass $Lm$.}
\end{figure}
\begin{figure}[htbp]
   \centerline{\epsfig{figure=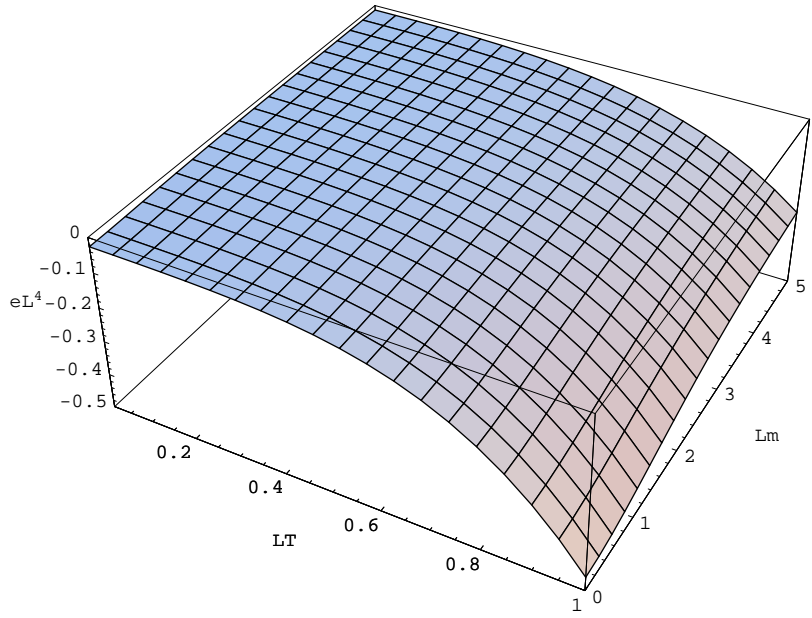,width=0.65\linewidth}}
\caption{The $T$ and $m$ dependence of Casimir energy $\epsilon_C(T) L^4$.}
\end{figure}
\begin{figure}[htbp]
   \centerline{\epsfig{figure=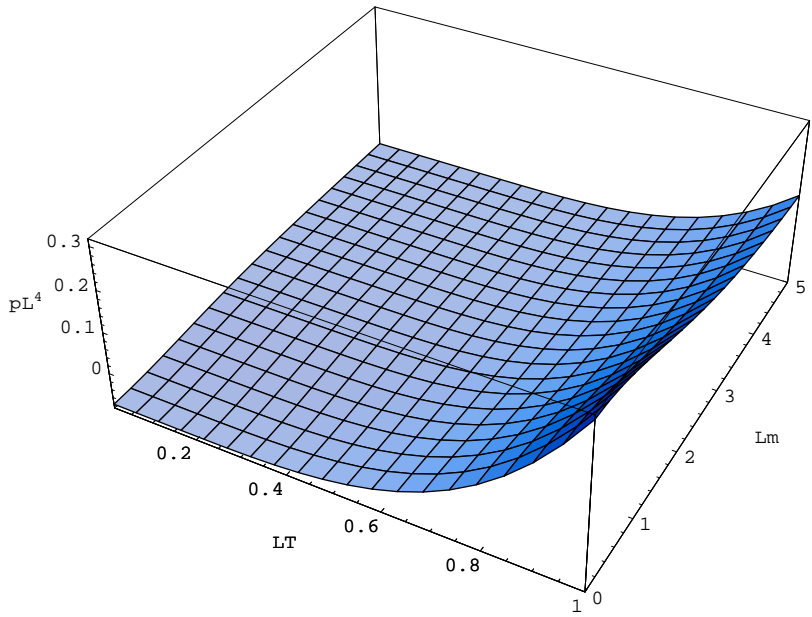,width=0.65\linewidth}}
\caption{The $T$ and $m$ dependence of Casimir pressure $p L^4$.}
\end{figure}

\end{document}